\documentclass[10pt,twocolumn,letterpaper]{article}

\usepackage{iccv}
\usepackage{times}
\usepackage{epsfig}
\usepackage{graphicx}
\usepackage{amsmath}
\usepackage{amssymb}
\usepackage{graphicx}
\usepackage{graphicx}
\usepackage{textcomp}
\usepackage{diagbox}
\usepackage{tabularx}
\usepackage{multirow}
\usepackage{capt-of} %
\usepackage{mathtools} %
\usepackage{color, colortbl}
\usepackage{bbding}
\usepackage{pifont}
\usepackage{wasysym}
\usepackage{amssymb}
\usepackage{arydshln}
\usepackage{placeins} 
\usepackage{adjustbox}
\usepackage{nicefrac}
\usepackage{booktabs}
\usepackage{amsmath} 
\usepackage{amsmath,amssymb,stmaryrd}   
\usepackage{colortbl}   
\usepackage{xcolor}
\usepackage{soul}
\usepackage[accsupp]{axessibility} 

\usepackage[pagebackref=true,breaklinks=true,letterpaper=true,colorlinks,bookmarks=false]{hyperref}

\iccvfinalcopy 

\def\iccvPaperID{4379} 

\ificcvfinal\fi

\begin{document}

\title{Preserving Tumor Volumes for Unsupervised Medical Image Registration}

\author{Qihua Dong$^1$\thanks{These authors contributed equally.}, Hao Du$^1$\textsuperscript{\textasteriskcentered}, Ying Song$^2$, Yan Xu$^3$\thanks{Corresponding authors} , Jing Liao$^1$\footnotemark[2] \\
$^1$Department of Computer Science, City University of Hong Kong\\
$^2$National Cancer Center/National Clinical Research Center for Cancer/Cancer Hospital, \\Chinese Academy of Medical Sciences and Peking Union Medical College\\
$^3$School of Biological Science and Medical Engineering, Beihang University\\
{\tt\small dongqh078@gmail.com, haodu8-c@my.cityu.edu.hk, jingliao@cityu.edu.hk}\\
{\tt\small songying1770@hotmail.com}\\
{\tt\small xuyan04@gmail.com}
}

\maketitle
\ificcvfinal\thispagestyle{empty}\fi

\begin{abstract}

Medical image registration is a critical task that estimates the spatial correspondence between pairs of images. However, current traditional and deep-learning-based methods rely on similarity measures to generate a deforming field, which often results in disproportionate volume changes in dissimilar regions, especially in tumor regions. These changes can significantly alter the tumor size and underlying anatomy, which limits the practical use of image registration in clinical diagnosis. To address this issue, we have formulated image registration with tumors as a constraint problem that preserves tumor volumes while maximizing image similarity in other normal regions. Our proposed strategy involves a two-stage process. 
In the first stage, we use similarity-based registration to identify potential tumor regions by their volume change, generating a soft tumor mask accordingly. In the second stage, we propose a volume-preserving registration with a novel adaptive volume-preserving loss that penalizes the change in size adaptively based on the masks calculated from the previous stage. 
Our approach balances image similarity and volume preservation in different regions, i.e., normal and tumor regions, by using soft tumor masks to adjust the imposition of volume-preserving loss on each one. This ensures that the tumor volume is preserved during the registration process. 
We have evaluated our strategy on various datasets and network architectures, demonstrating that our method successfully preserves tumor volume while achieving comparable registration results with state-of-the-art methods. Our codes is available at: \url{https://dddraxxx.github.io/Volume-Preserving-Registration/}.
\end{abstract}

\section{Introduction}

\begin{figure}[htbp]
\begin{center}
    \includegraphics[width=0.4\linewidth]{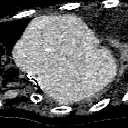}
    \makebox[0.05\linewidth]{}
    \includegraphics[width=0.4\linewidth]{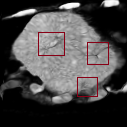}
    \\
    \makebox[0.4\linewidth]{\scriptsize Fixed Image}
    \makebox[0.05\linewidth]{}
    \makebox[0.4\linewidth]{\scriptsize Similarity-Based Registration}
    \\
    \includegraphics[width=0.4\linewidth]{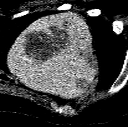}
    \makebox[0.05\linewidth]{}
    \includegraphics[width=0.4\linewidth]{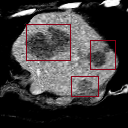}
    \\
    \makebox[0.4\linewidth]{\scriptsize Moving Image}
    \makebox[0.05\linewidth]{}
    \makebox[0.4\linewidth]{\scriptsize Volume-Preserving Registration}
\end{center}
\caption{Red boxes represent the location of tumors in the moving image before warped. In image registration with tumors, similarity-based registration typically leads to significant alterations in tumor size while our volume-preserving registration is capable of preserving tumor anatomy.}
\label{fig:Fig1}
\vspace{-0.5cm}
\end{figure}

Deformable image registration is a fundamental task that estimates non-linear spatial correspondences between two images. It is useful for medical image studies that involve 3D images of organs or tissues, such as MR brain scans and CT liver scans. 
Recently, a wide range of deep-learning-based methods are proposed in the field, with better performance and obvious speed-ups in inference time compared to the traditional registration methods ~\cite{balakrishnan2018unsupervised, zhao2019recursive, wang2022transformer, hoffmann2021synthmorph, hoffmann2023anatomy}. 

Currently, most of the learning-based methods~\cite{balakrishnan2018unsupervised, zhao2019recursive} train a registration model to achieve high similarity in either image intensity~\cite{zhao2019recursive, balakrishnan2018unsupervised} or anatomy label maps~\cite{hoffmann2021synthmorph}. 
While this is useful in various medical applications, such as atlas-based segmentation and image fusion~\cite{chen2022joint}, problem emerges, however, when studying images with tumors~\cite{baheti2021brain,elazab2018macroscopic, mok2022unsupervised}. 
Tracking tumor growth is a core task in cancer treatment, which can be used to evaluate the outcomes of radiotherapy and chemotherapy, and plan optimal postoperative treatment~\cite{elazab2018macroscopic}. The process requires registration to align the anatomy of images from different periods of the patient, while preserving tumor properties~\eg, the size. 
Traditional models using regular registration have struggled to accomplish this task~\cite{elazab2018macroscopic}. We have observed that this remains an issue for current learning-based methods, and a viable solution has yet to be proposed. 

As shown in Figure \ref{fig:Fig1}, the volume of tumors significantly reduces when using popular unsupervised registration networks to align tumor images with atlas images. The issue is widespread since mainstream registration networks focus on optimizing the similarity between image pairs while ignoring tumor regions. Tumor regions often lack corresponding parts, and even for images of the same patient, the size and location of tumors may vary greatly. This can happen because tumors can change shape, size, location, or disappear over time. 
Therefore, in deformable settings, the size of tumors will change disproportionately to maximize the image similarity. This problem may be more severe when the registration network has better warping ability, leading to better performance in evaluations from previous works, as confirmed by our experiments in Section \ref{Exp}.
The disproportionate change of tumor volume in deformable registration is lethal for clinical evaluation of tumor growth and greatly hinders the application of registration in clinics.

In this paper, we re-formulate the deformable registration problem as a constraint problem that preserves the properties of tumors in the original images, \ie, the shape and the size of the tumor, while maximizing image similarity in non-tumor regions. For instance, in tumor image registration with atlas, besides the alignment of anatomical structures, the tumor should have similar morphing behaviours as its surrounding organs, \ie, the size of it should change proportionally to the size of the organ. 
However, this poses two major challenges for our strategy. Firstly, due to the limited availability of annotated data, most existing registration methods resort to unsupervised learning. Therefore, a key question is how to identify tumor regions in the unsupervised setting. 
Second, after the tumor regions have been identified, the challenge still remains of preserving tumor volumes while simultaneously maximizing similarity in other regions. Finding a way to balance these two competing objectives is crucial for achieving effective tumor image registration.


To tackle the challenges, we propose a novel two-stage strategy that operates in an unsupervised manner. In the first stage, we leverage a similarity-based registration network to identify regions that undergo excessive volume change during registration. Such regions are indicative of the presence of tumors, as tumors tend to exhibit larger changes in volume due to a lack of correspondence between image pairs. The outcome of the first stage is a soft tumor mask that distinguishes normal and possible tumor regions.
In the second stage, we introduce an adaptive volume-preserving loss to train a volume-preserving registration network. Based on the soft tumor mask obtained from the first stage, this loss is designed to adaptively penalize volume changes in different regions of the image. Possible tumor regions are assigned larger loss weights, and similarity loss weights are adjusted accordingly in the opposite direction. The application of the adaptive volume-preserving loss effectively balances similarity and volume preservation in different parts of the image.

It should be noted that the soft masks from stage one do not need to be highly accurate to achieve desirable results, as our stage two is robust, which is demonstrated by our ablation experiments in section \ref{Exp}. In addition to evaluating the robustness of our proposed strategy on various datasets and frameworks, including CNN and transformer, as well as different types of imaging data such as brain and abdominal, MRI and CT images, we have also introduced the Square Tumor Size Ratio (STSR) metric to measure the preservation of tumor volumes. The results demonstrate that our methods achieve comparable warping performance metrics such as Dice coefficient and average landmark distances while effectively preserving tumor properties as measured by the STSR.

To conclude, our contribution includes: 
(1) We have reformulated image registration with tumors as a constraint problem that preserves tumor volumes while maximizes image similarity in other normal regions.
(2) We have designed a two-stage process that the first stage performs regular registration to estimate tumor regions, and the second stage performs volume-preserving registration on tumor regions based on the adaptive volume-preserving loss.
(3) We have proposed the STSR metric to measure the preservation of tumor volumes in registration, and have evaluated the effectiveness of our strategy on different network architectures and various datasets.

\section{Related Work}

\subsection{Learning-Based Deformable Image Registration}

Compared to conventional registration methods, deep networks have achieved remarkable registration accuracy and speed using various network structures, \ie, CNN and Transformer, in  applications like atlas-image registration~\cite{shi2022xmorpher,chen2022deformer,kim2021cyclemorph,chen2022transmorph,wang2022transformer,zhao2019recursive,balakrishnan2018unsupervised, hoffmann2021synthmorph, hoffmann2023anatomy, chen2022joint}. They generally adopts unsupervised learning, utilizing image similarity on either image intensities~\cite{shi2022xmorpher, chen2022deformer, chen2022transmorph, zhao2019recursive} or synthesized label maps~\cite{hoffmann2021synthmorph, hoffmann2023anatomy} as well as smoothness regularization to supervise the registration network~\cite{chen2022deformer, chen2022transmorph, hoffmann2021synthmorph,hoffmann2023anatomy,zhao2019recursive}. Advancements are often made through the introduction of novel regularization such as cycle consistency~\cite{kim2021cyclemorph} and geometry preservation~\cite{arar2020unsupervised}, or through innovative network structures like multi-scale architectures~\cite{shao2022multi, mok2020large}. 
However, existing methods do not adequately consider cases where tumors exist. Tumors often have no corresponding part between image pairs, and therefore, optimizing for similarity can excessively change their volumes, which limits the application of registration networks in critical clinical applications like cancer treatment. To the best of our knowledge, we are the first to study this problem with learning-based methods and propose a metric to measure volume preservation specifically for tumors.

\subsection{Deformable Image Registration with Tumors}
Conventional registration models often resolve registering images with tumor by (1) excluding the similarity measure on tumors~\cite{OlivierClatz2005RobustNR, DongjinKwon2014PORTRPA}, which requires either massive manual segmentation~\cite{OlivierClatz2005RobustNR} or initial seed to explicitly model tumor-growth~\cite{elazab2018macroscopic, DongjinKwon2014PORTRPA}. 
(2) jointly segmenting and registering images~\cite{NichaChitphakdithai2010NonrigidRW}, which is time-consuming since it need to iteratively detect dissimilarity in image intensities and perform resection and retraction. 
(3) reconstructing quasi-normal images from pathological ones~\cite{XuHan2017EfficientRO,DongjinKwon2015EstimatingPS}, which is also slow since the reconstruction need to be iteratively improved. 

Recently, learning-based methods for registration with tumors~\cite{mok2022unsupervised, meng2022brain, wodzinski2022unsupervised} are proposed for the release of BraTS-Reg dataset~\cite{baheti2021brain}. They aim to improve the spatial correspondence for non-tumor regions, thus either directly annotates landmarks to supervise registration~\cite{meng2022brain} or introducing invertibility of the registration to mask non-inverted regions and ensure high correspondence in others~\cite{mok2022unsupervised, wodzinski2022unsupervised}.
However, these methods assume the absence of tumors in one image of the registered pair, which may not be applicable in practical applications where we need to compare two tumor images from one patient to track the tumor growth. In this scenario, invertibility in tumors is still possible and thus, it is essential to consider the preservation of tumor properties during registration.

\subsection{Unsupervised Tumor Segmentation}
To ensure the preservation of tumor volumes during registration, it is crucial to first identify the locations of tumors. Current deep learning methods have made remarkable progress in unsupervised tumor segmentation, as demonstrated by recent studies~\cite{yao2021label, zhang2023self, lyu2022learning}. These methods typically use GAN models or simulate tumors on healthy images to generate synthetic data, which is then used to train the segmentation network. Although the well-trained network can achieve high segmentation accuracy, it often requires complex training strategies and additional models or testing steps to achieve the desired performance, as noted in recent research~\cite{zhang2023self, lyu2022learning}. However, in our approach, we do not rely on sophisticated segmentation methods to accurately segment tumors. Instead, our experiments, as presented in Section~\ref{Exp}, demonstrate that our tumor mask estimation is sufficient for our strategy to learn a volume-preserving registration for tumors.

\subsection{Volume-Preserving Deformable Image Registration}
Volume-preserving registration has been studied extensively to improve the accuracy of anatomical registration. In pulmonary CT image registration, for example, tissue intensity changes between inspiration and expiration phases can be utilized as prior knowledge, and adding a volume-preserving constraint has been shown to be effective~\cite{DiWang2022PLOSLPL,Zhao2016TissueVolumePD}. In other cases, such as when soft tissues are incompressible or when images come from different modalities, a volume-preserving constraint can also serve as a helpful regularization~\cite{rohlfing2003alternating, ChristineTanner2002ValidationOV}.
Our work, instead, focus on using volume-preservation to maintain the properties of tumors during image registration. 
The proposed strategy imposes this constraint adaptively on different parts of the images in order to preserve tumor volumes and ensure anatomical accuracy, while previous works have either preserved volumes for the entire images~\cite{rohlfing2003alternating, ChristineTanner2002ValidationOV} or can use simple methods to determine the regions to be preserved~\cite{DiWang2022PLOSLPL, Zhao2016TissueVolumePD}.

\begin{figure*}[t]
\begin{center}
   \includegraphics[width=0.9\linewidth]{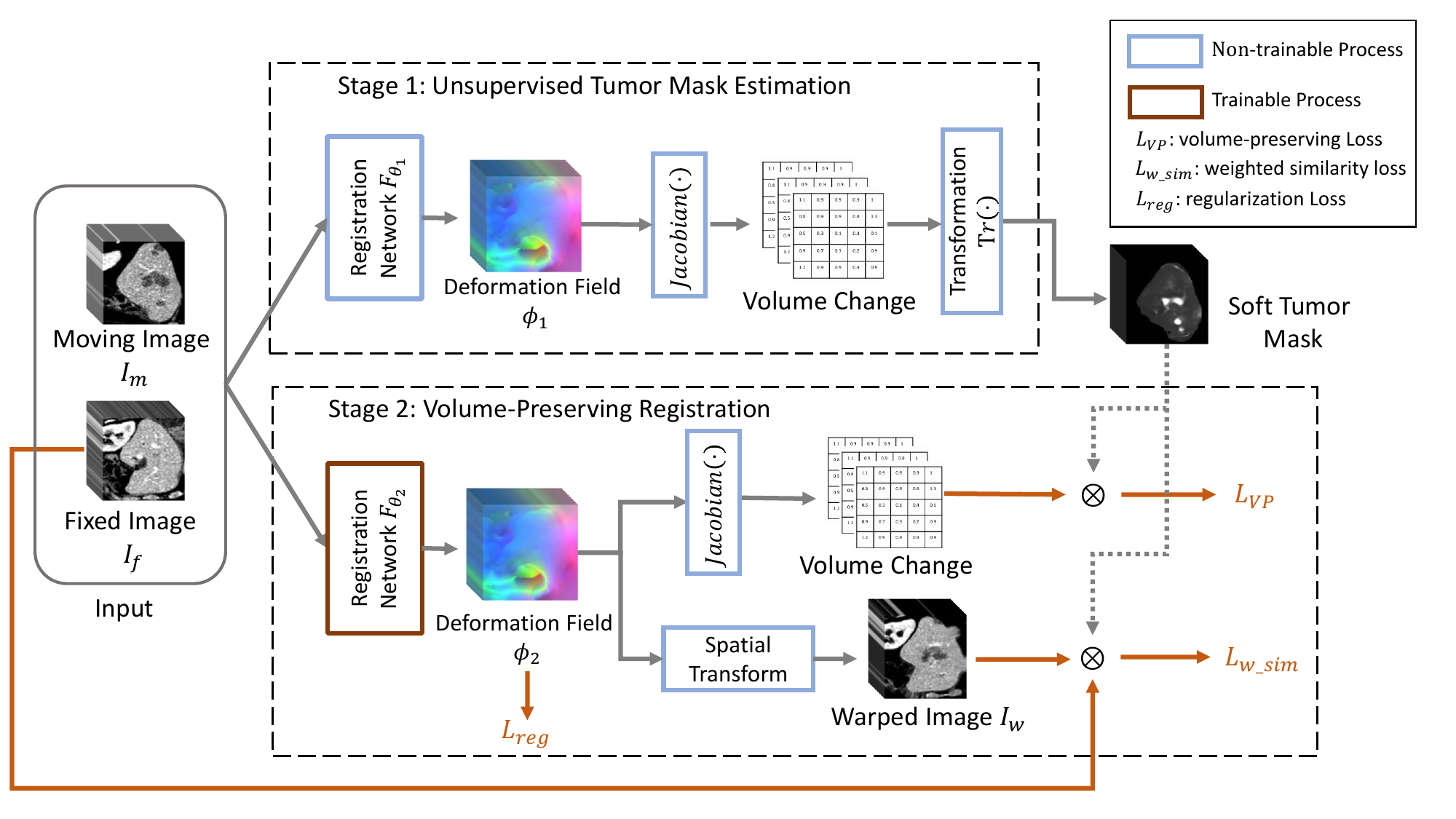}
\end{center}
\vspace{-0.5cm}
   \caption{
   Main Framework. Our strategy consists of two stages. In the first stage, a soft mask indicating tumor regions is estimated by analyzing the Jacobian matrix of the deformation field obtained by a registration network pre-trained on similarity loss. In the second stage, the soft mask is used to guide the calculation of both volume-preserving and similarity losses for training the volume-preserving registration network. This deformation field estimated by the stage two ensures that both the image similarity and the preservation of tumor sizes.
   }
\label{fig:Fig2}
\vspace{-0.2cm}
\end{figure*}

\section{Methodology}

\subsection{Preliminaries}
\textbf{Problem Setting.} Given a moving image $I_m$ and fixed image $I_f$ defined over $d$-{dimensional} (d=3 in this paper) space $\Omega$, deformable image registration aims to find a dense deformation filed $\phi:\Omega\to\Omega$. This field warps $I_m$, meaning it projects every pixel position $x$ in warped image $I_w$ to $I_m$, to achieve alignment with $I_f$. 
In this paper, the task is to construct a registration network $F_\theta$ that takes input $I_m, I_f$ to predict the deformation field $\phi$. 
\begin{align}
    \phi &= F_\theta(I_m, I_f)
    \\
    I_w(x) &= I_m(\phi(x)+x)
    \\
    I_w &\sim I_f
\end{align}

\textbf{Evaluation Criteria.} In general, a successful deformable image registration involves accurately aligning the anatomical structures while also ensuring a smooth deformation field. This is often achieved through the use of similarity-based objectives and regularization methods that promote smoothness. However, in the context of image registration with tumors, simply aligning the anatomy is not sufficient. Preserving the volume of the tumors is crucial for accurately tracking tumor growth, as previously discussed.
To be specific, given the moving image $I_m$ with the tumor $T_m$ and its surrounding organ $O_m$ and the warped moving image $I_w$ with $T_w$ and $O_w$, the morphing of tumor $T_m$ should be similar to that of $O_m$. Thus in respect of the volume preservation, the change of tumor size $|T_m|$ should be proportionate to the change 
of the containing organ $|O_m|$, \ie: 
\begin{align}
    TSR(I_m) &= {|T_m|\over |O_m|} \label{eq:tsr}\\
   TSR(I_m) &\approx TSR(I_w)
   \label{eq:vol-pre}
\end{align}
We therefore define the tumor size ratio (TSR) as $|T|\over |O|$, which should be preserved after registration, and the re-formulated registration objective is now maximizing image similarity while preserving the volumes of tumors, \ie, the TSR.

\subsection{Overall Strategy}
The proposed strategy consists of two stages, as shown in Figure~\ref{fig:Fig2}. The first stage estimates potential tumor regions and generates a soft mask for them. This is done by analyzing volume changes, specifically the Jacobian matrix, in the deformation field obtained by an existing similarity-based registration network. As tumor regions tend to undergo excessive volume change during registration, they can be identified through this analysis. In the second stage, we perform volume-preserving registration for tumors using an adaptive volume-preserving loss. This loss is designed to guide the volume-preserving process effectively and is associated with the tumor masks estimated in the first stage. 
Through the assignment of distinct weights to diverse regions, including both normal and tumor regions, it is possible to preserve tumor volumes while also ensuring a high degree of similarity in other regions.

Importantly, our strategy can be applied to nearly any learning-based registration network without introducing any additional modules.


\subsubsection{Stage 1: Unsupervised Tumor Mask Estimation}
Following previous works~\cite{balakrishnan2018unsupervised, chen2022transmorph, zhao2019recursive}, we first train a similarity-based registration network $F_{\theta_1}$ with parameters $\theta_1$ and fix its parameters. As shown in Figure~\ref{fig:Fig1}, tumors in medical images can be detected by analyzing the change in volume through similarity-based registration. Our goal is to utilize this characteristic to facilitate the unsupervised estimation of the tumor mask for volume preservation during registration.

To determine the volume change at voxel $x$, we calculate the determinant of the Jacobian matrix of the deformation field ($J_{\theta}$) with model parameter $\theta$. Our objective is to ensure that the volume change in each point of the tumor region is similar to the volume change of the organ, which is an indicator of the organ's size change (as shown in Equation~\ref{eq:vol-tr}). To measure the distance between the two, we calculate the number of times one ratio exceeds another, denoted as $D_\theta$, using Equations~\ref{eq:dist} and~\ref{eq:times} where $D_{\theta}'(x)$ represents the relative volume change of voxel $x$ with respect to organ areas.
\begin{align}
     {\frac{|T_w|}{|O_w|}} / \frac{|T_m|}{|O_m|} \approx 1 &\Rightarrow {\frac{|T_w|}{|T_m|}} / \frac{|O_w|}{|O_m|} \approx 1, \text{from Eq.~\ref{eq:vol-pre}}, \label{eq:vol-tr} \\
     D_{\theta}'(x) &= {|J_{\theta(x)}|} / {\frac{|O_w|}{|O_m|}}, x \in \Omega ,\label{eq:times} \\
     D_{\theta}(x) &= max(D_{\theta}'(x), 1/D_{\theta}'(x)),\label{eq:dist}
\end{align}

In practice, we estimate the organ masks $O_m$ and $O_w$ by warping a reference image with the ground truth (GT) organ segmentation. We randomly select the reference image from the training dataset, and the choice of reference image does not significantly impact the results because registration for organs achieves high accuracy, as evidenced by the dice value in table ~\ref{tab:tab1}. By warping the reference image, we obtain the organ segmentation for the moving image $O_m$. We then calculate the organ segmentation for the warped image $O_w$.

To obtain a soft tumor mask $STM$ from the distance $D_{\theta_1}$ defined in~\ref{eq:dist}, we use a transformation function $Tr$. This function can be formulated as the following equation:
\begin{align}
    STM(x) &= Tr(D_{\theta_1}(x)), \label{eq:STM}
\end{align}
{Here, $D_{\theta_1}(x)$ represents the distance value at pixel $x$, and $Tr$ transforms values from the range of $[0,\infty]$ to the range of $[0,1]$. A value of 1 indicates the presence of a tumor, while a value of 0 indicates a normal region. In practical implementation, we use $Tr(x) = Sigm(5\cdot(x-1.5))$ since the sigmoid function is widely adopted for converting values to the range of $[0,1]$ and to our experience, the size of tumor regions averagely changes around 1.5 times. However, the choice of transformation functions is not exclusive, as demonstrated in our experiments in Section~\ref{Exp}. We also set the regions outside the organ mask to $0$.}

\begin{figure}[]
\begin{center}
    \includegraphics[width=0.4\linewidth]{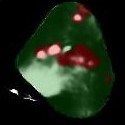}
    \makebox[0.05\linewidth]{}
    \includegraphics[width=0.4\linewidth]{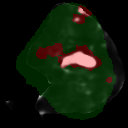}
    \\
    \makebox[0.45\linewidth]{ \scriptsize (a)Tumor Mask}
    \makebox[0.05\linewidth]{}
    \makebox[0.45\linewidth]{ \scriptsize (b)Tumor Mask after pre-registration}
\end{center}
\caption{Visualization of the soft tumor mask. (a) is the estimated soft tumor mask without pre-registration and (b) is the one with pre-registration. The gray colors with intensity variations denote the soft mask, while the green and red colors represent the ground truth segmentation for the liver and tumors, respectively.}
\label{fig:Fig3}
\vspace{-0.35cm}
\end{figure}

We observed that some normal regions, particularly those close to the organ boundary, may also exhibit some volume changes after registration. As a result, these regions may be incorrectly included in the tumor mask, as shown in Figure~\ref{fig:Fig3} (a). To address this issue, we implemented a pre-registration step to first align the edges and exclude volume changes caused by edge alignment.  Specifically, in the pre-registration step, we applied a \textit{bilateral filter} to the moving image, which preserves the edges while smoothing the interior regions. We then used the registration network to register the filtered moving image with the fixed image in order to align their edges. After this, an unfiltered warped image was obtained by spatially transforming the unfiltered moving image with the deformation field. The second registration process was then performed on the unfiltered warped image, which focused on the registration of interior regions and was able to accurately estimate a tumor mask, as shown in Figure~\ref{fig:Fig3} (b). Therefore, this pre-registration approach effectively selects the tumor regions while filtering out the non-tumor regions.


\subsubsection{Stage 2: Volume-Preserving Registration for Tumors}

After obtaining the soft tumor mask $STM$, we can adaptively select tumor regions to maintain the size ratio. Here, we define the volume-preserving loss directly as the distance $D_{\theta_2}$, as shown in Equation \ref{eq:dist}. By minimizing the distance, we can achieve smaller volume changes, indicating better preservation of tumor volume.
The volume-preserving loss is multiplied by the soft tumor mask. This ensures that the volume-preserving constraints are adaptively applied to the tumor regions and not to the normal regions, {which can be formulated as:}
\begin{align}
L_\text{VP} &= {\frac{1}{|I_m|}}\sum_{x\in \Omega}D_{\theta_2}(x) \cdot STM(x),
\end{align}
{Here, in the second stage}, the registration network takes the soft tumor mask as input, which serves as an indicator for the regions that need to be preserved in terms of volume. 
We follow the similarity and registration loss formulation in~\cite{zhao2019recursive}, but we use a weighted similarity loss to adapt to the volume-preserving loss, where $Cov$ is the covariance between two images $I_1, I_2$, $\hat{I}$ is the mean of $I$, and $Cov'$ is the adapted version utilizing $STM$ weight:

\begin{align}
\operatorname{Cov'}\left[I_1,I_2\right]&={\frac{\sum(I_{1}-\hat{I_1})(I_{2}-\hat{I_2})(1-STM)}{\sum(1-STM)}},
\\
    L_\text{w\_sim}(x) &= \frac{\text{Cov'}\left[I_w,I_f\right]}{\sqrt{\text{Cov} [ I_w,I_w ] \text{Cov} \left[I_f,I_f \right]}},
\end{align}
{Therefore, the training loss of the second stage is}: 
\begin{align}
L = L_\text{w\_sim} + \alpha_1 \cdot L_\text{VP} + \alpha_2 \cdot L_\text{reg},
\end{align}
where we use the smoothness regularization from VoxelMorph~\cite{balakrishnan2018unsupervised} as $L_{reg}$, and $\alpha_1$ and $\alpha_2$ as hyperparameters to control the relative importance of the volume-preserving loss and the regularization loss.


\section{Experiments}
\label{Exp}

\subsection{Experimental Settings}

\noindent\textbf{Implementation.}
Following the implementation of Recursive Cascaded Network (RCN)~\cite{zhao2019recursive}, we employed PyTorch and utilized the same similarity loss and regularization losses (for affine and deformable, respectively) for baseline methods. The weight of the volume-preserving losses is set to 0.1 while the weights of similarity loss is 1 and regularization loss is 0.1. Models were trained on one NVIDIA GeForce RTX 3090 GPU with a batch size of 4. During the training stage, we used the Adam optimizer~\cite{zhang2018improved} and ran for a total of 5 epochs, with 20000 iterations in each epoch. The learning rate was set to $10^{-4}$.
\\
\\

\begin{figure*}[htbp]
\begin{center}
   \includegraphics[trim={0 150 0 140},clip,width=\linewidth]{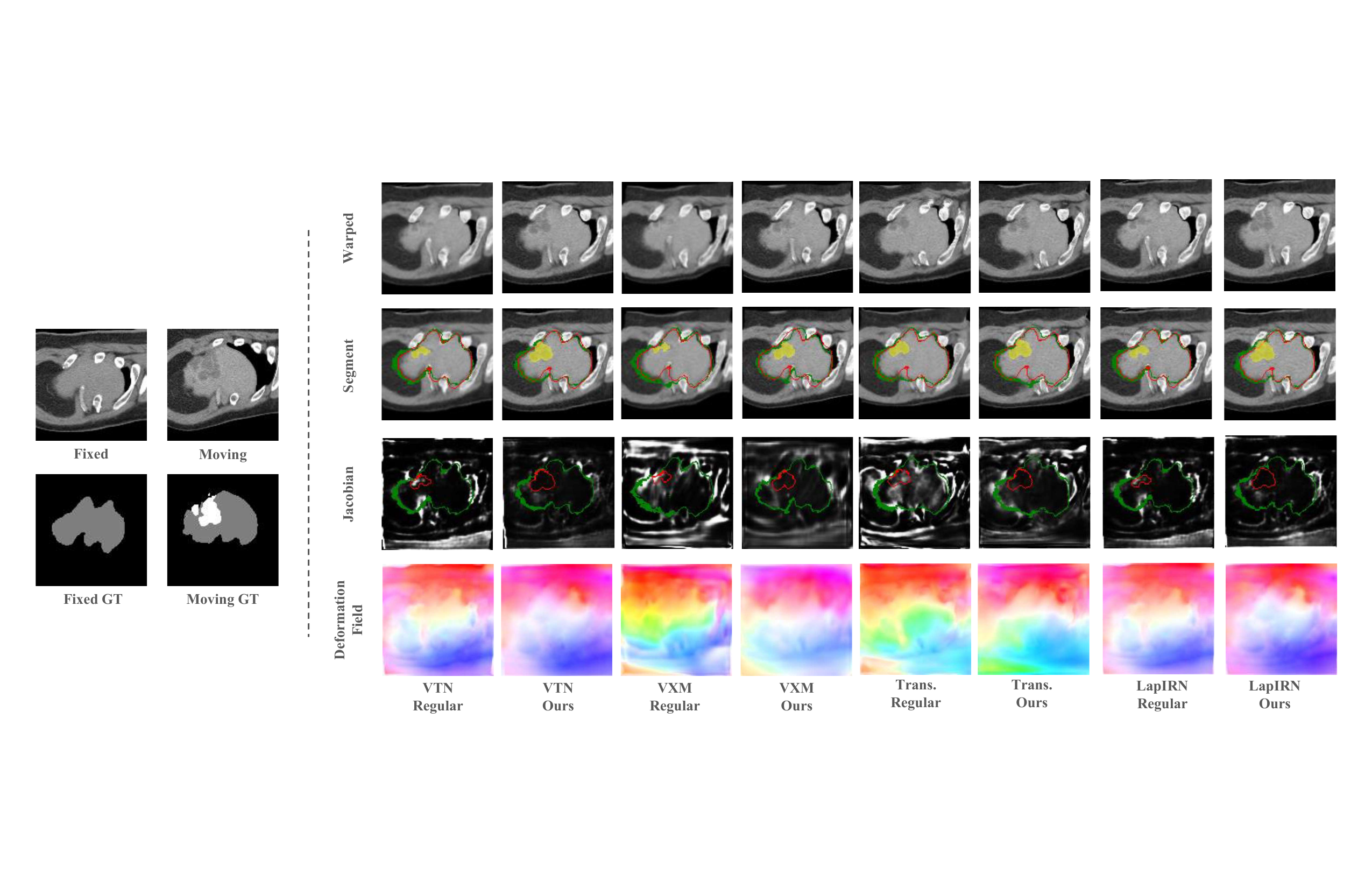}
\end{center}
   \caption{Qualitative comparison of different volume-preserving (VP) methods trained on the Liver Tumor Segmentation (LiTS) dataset. The left side of the figure shows two sets of images: Fixed and Ground Truth (GT), and Moving and GT.
The first row of the figure displays the warped moving image, while the second row illustrates the organ outlines in green and red for the moving and fixed images, respectively. The yellow overlay highlights the tumors. Our proposed volume-preserving (VP) method ensures the preservation of tumor volume while aligning the images, as demonstrated by reduced number of visible changes in tumor size.
In the third row, the Jacobian matrix of the deformation field is visualized. The green and red lines represent the organ and tumor outlines, respectively. The white areas indicate a large Jacobian, which corresponds to a more significant change in volume. The method without volume-preserving loss demonstrates a larger white area in the tumor, indicating a greater volume change of tumor volume.
The last row of the figure displays the deformation field. 
Due to space limitations, qualitative results for BraTS20 dataset are provided in the supplemental material.
   }
\label{fig:Fig4}
\end{figure*}

\newcommand{\setgray}{\cellcolor[gray]{.9}}
\begin{table*}[]
\centering
\resizebox{1.\linewidth}{!}{
\begin{tabular}{lc ccccc c ccccc}
\specialrule{.8pt}{0pt}{2pt}
\multirow{2}{*}{Network}&\multirow{2}{*}{Methods}&\multicolumn{5}{c}{LiTs17}    &&\multicolumn{5}{c}{BraTS20}                                                                                            \\ 
\cline{3-7} \cline{9-13}
    &  & Dice $\uparrow$ & Lm. Dist $\downarrow$ & $|J_\theta|<0$(\%) $\downarrow$  & Std. $|J_\theta|$ $\downarrow$ & STSR $\downarrow$  &&   Dice $\uparrow$ & Lm. Dist $\downarrow$ & $|J_\theta|<0$(\%) $\downarrow$  & Std. $|J_\theta|$ $\downarrow$ & STSR $\downarrow$      \\ \specialrule{.4pt}{2pt}{0pt}
SyN        & -        & 0.850  & 12.36    & 0.192               & 0.221             & 1.564     && 0.9560 & 4.01     & 0.276               & 0.226             & 1.512      \\
B-Spline   & -        & 0.853  & 14.22    & 0.003               & 0.288             & 1.330  && 0.9601 & 4.51     & 0.001               & 0.171             & 1.622         \\ 
\specialrule{.4pt}{2pt}{0pt}

& \multicolumn{1}{l}{Regular} & 0.912  & 10.46    & 3.545               & 2.048             & 2.330     && 0.9784 & 2.98     & 0.396               & 0.725             & 2.263     \\
\multirow{-2}{*}{VTN}           & \setgray Ours   & \setgray0.908  & \setgray10.77    & \setgray2.956               & \setgray1.124             & \setgray\textbf{1.260} &\setgray& \setgray0.9738 & \setgray3.10     & \setgray0.596               & \setgray0.285             & \setgray\textbf{1.416}\\ 
\specialrule{.4pt}{2pt}{0pt}

        & \multicolumn{1}{l}{Regular} & 0.860  & 12.62    & 4.191               & 1.699             & 1.738     && 0.9701 & 3.56     & 0.024               & 0.448             & 1.929      \\
           \multirow{-2}{*}{VXM} & \setgray Ours    & \setgray0.857  & \setgray12.43    & \setgray1.681               & \setgray0.691             & \setgray\textbf{1.241} &\setgray& \setgray0.9741 & \setgray3.60     & \setgray0.030               & \setgray0.226             & \setgray\textbf{1.463}\\ 
           \specialrule{.4pt}{2pt}{0pt}
  
 & \multicolumn{1}{l}{Regular} & 0.859  & 13.35    & 6.377               & 1.906             & 1.785    && 0.9710 & 4.01     & 0.164               & 0.468             & 1.868             \\
           \multirow{-2}{*}{TransMorph}& \setgray Ours    & \setgray0.856  & \setgray13.19    & \setgray4.084               & \setgray0.827             & \setgray\textbf{1.302} &\setgray& \setgray0.9758 & \setgray3.98     & \setgray0.019               & \setgray0.194             & \setgray\textbf{1.319}\\ 
\specialrule{.4pt}{2pt}{0pt}    

 & \multicolumn{1}{l}{Regular} & 0.8933  & 10.23   & 5.301                & 1.788              & 2.360    &&  0.9750 & 3.40 & 3.020 & 0.301 & 1.889           \\
\multirow{-2}{*}{LapIRN}& \setgray Ours    & \setgray 0.8893 & \setgray 10.55   & \setgray 4.120               & \setgray 0.988             & \setgray\textbf{1.645} &\setgray& \setgray 0.9743 & \setgray 3.36 & \setgray 0.040 & \setgray 0.105 & \setgray \textbf{1.224}\\ 
\specialrule{.8pt}{0pt}{2pt}
\end{tabular}
}
\caption{Quantitative comparison with state-of-the-art registration methods. 
Our strategy was applied to two datasets, namely LiTS17 and BraTS20, and three network architectures
In addition, five metrics were employed for the evaluation process.
The results of the comparison indicate that our proposed strategy can be well generalized to other state-of-the-art methods to preserve tumor volumes while achieving comparable or better performance on other metrics. 
}
\label{tab:tab1}
\end{table*}

\noindent\textbf{{Baselines}.}
We re-implemented our strategy on {four} SOTA learning-based models that have distinct network architectures: {1) recursive cascaded network with VTN base network (VTN)~\cite{zhao2019recursive}, which stacks CNN networks to achieve higher registration accuracy; 2) VoxelMorph~\cite{balakrishnan2018unsupervised}, which is a widely used CNN network that utilize stationary vector field to register images; 3) TransMorph~\cite{chen2022transmorph}, which is a well-performed hybrid Transformer-ConvNet network; 4) LapIRN~\cite{mok2020large}, which is a Laplacian Pyramid network designed to operate in a coarse-to-fine manner. We fit these learning-based networks into our strategy with minor efforts and achieve good performance. Additionally, we compare our proposed method with two state-of-the-art traditional methods for deformable image registration, SyN~\cite{avants2008symmetric} (integrated in ANTs~\cite{avants2009advanced} with the affine stage) and B-spline~\cite{rueckert1999nonrigid} (integrated in Elastix~\cite{klein2009elastix} with the affine stage) following~\cite{zhao2019recursive}. }

\noindent\textbf{Evaluation Metrics.}
{The evaluation of the proposed strategy is composed of five metrics: Dice score, landmark distance to measure the anatomical alignment, 
and Square Tumor Size Ratio metric to measure the volume preservation of tumors.} 
More specifically, the Dice score is calculated using the following formula:
\begin{align}
Dice(I_1, I_2) = {2 \cdot |I_1 \cap I_2| \over |I_1| + |I_2|},
\end{align}
Additionally, we also use landmark annotations to measure the anatomical alignment between the warped image and the fixed image. We calculate the average distance between the fixed image's landmarks and the warped landmarks of the moving image, as also introduced in RCN~\cite{zhao2019recursive}. 
To quantify the diffeomorphism and smoothness of the deformation fields, we followed the methodology of previous works such as~\cite{chen2022deformer, chen2022transmorph, zhao2019unsupervised}. Here, two metrics were employed: the average percentage of voxels with non-positive Jacobian determinant ($|J_\theta|<0(\%)$) in the deformation fields, and the standard deviation of the Jacobian determinant ($Std. |J_\theta|$). 

To measure the volume preservation of tumors, we propose Square Tumor Size Ratio metric (STSR), which is calculated from the change of tumor size ratio (TSR) defined in Equation~\ref{eq:tsr}:
\begin{align}
    STSR &=  max(({TSR(I_m)\over TSR(I_w)}),({TSR(I_w)\over TSR(I_m)}))^2,
\end{align}
Here, $I_m, I_w$ refers to the moving image and the warped moving image. 
The square term is used o account for larger changes in tumor volumes, which can have a more significant impact on the accuracy of tumor growth estimation in clinical applications. By utilizing the STSR metric, we are able to accurately evaluate the performance of our proposed strategy.

\noindent\textbf{Preprocessing.} In our study, we follow standard preprocessing steps as outlined in VTN~\cite{zhao2019unsupervised} and VoxelMorph~\cite{balakrishnan2018unsupervised}. The raw scans are resampled into 128 × 128 × 128 voxels after cropping any unnecessary areas around the target object. 

\subsection{Datasets}
 \noindent\textbf{Training.} To train our model, we utilized two publicly available datasets: (1) LiTS17~\cite{lits}: The LiTS dataset comprises 201 CT scans that were acquired using various CT scanners and devices. The resolution of the images in this dataset ranges from $0.56$mm to $1.0$mm in the axial direction and from 0.45mm to 6.0mm in the $z$ direction. The number of slices in the $z$ direction ranges from 42 to 1026.
(2) BraTS20~\cite{brats2020}: It consists of 369 MRI scans of brain tumors, each with ground truth segmentation of four different tumor components. For training our neural network, we utilized the T1ce modality and treated all four components as tumors, while considering the entire brain as the surrounding organ.

\noindent\textbf{Testing.} For testing, we utilized a set of 10 cases of CT liver scans with metastatic tumors, and 10 MRI brain scans with metastatic tumors. The modality of MRI brain scans is T1ce.
Each case included both pre- and post-treatment images of the same patients, resulting in a total of 20 scans for each set. 
Following~\cite{shan2017unsupervised}, the segmentation of the scans, which contains the organ (liver or brain) and the tumor inside, were carefully annotated by three experts in the field. 
Landmarks were also annotated by three experts in the field to evaluate the registration accuracy of our strategy. 


\begin{table*}[htbp]
\centering
\resizebox{.95\linewidth}{!}{
\begin{tabular}{ccccc ccccc} 
\hline
\multirow{2}{*}{\#}&\multirow{2}{*}{Network} & \multirow{2}{*}{Unsupervised} &\multirow{2}{*}{Adaptive} & \multirow{2}{*}{Region} &  \multicolumn{5}{c}{LiTs17} \\
\cline{6-10}
& & &  &  & Dice $\uparrow$ & Lm. Dist $\downarrow$ & $|J_\theta|<0$(\%) $\downarrow$  & Std. $|J_\theta|$ $\downarrow$ & STSR $\downarrow$ \\
\hline
1&\multirow{5}{*}{VTN}                  & \checkmark           &     \textendash     & \multicolumn{1}{c|}{\textendash}      & 0.912                    & 10.46                  & 3.545                   & 2.048                    & 2.330                    \\
2&& \texttimes           & \textendash          & \multicolumn{1}{c|}{\textendash}      & 0.914                    & 10.22                  & 3.241                   & 1.165                    & 1.648                    \\
3& & \texttimes           & \texttimes           & \multicolumn{1}{c|}{Tumor} & 0.911                    & 10.74                  & 2.883                   & 1.017                    & 1.205                    \\
4&& \checkmark           & \texttimes           & \multicolumn{1}{c|}{Organ} & 0.904                    & 11.07                  & 3.194                   & 1.015                    & 1.640                    \\
5&& \setgray\checkmark           & \setgray\checkmark           & \multicolumn{1}{c|}{\setgray Organ} & \setgray0.908                    & \setgray10.77                  & \setgray2.956                   & \setgray1.124 & \setgray1.260 \\
\hline 
6&\multirow{5}{*}{VXM}                      & \checkmark           &    \textendash                 &  \multicolumn{1}{c|}{\textendash}      & 0.860                    & 12.62                  & 4.191                   & 1.699                    & 1.738                    \\
7& & \texttimes           & \textendash          & \multicolumn{1}{c|}{\textendash}      & 0.863                    & 12.22                  & 1.603                   & 0.691                    & 1.417                    \\
8&& \texttimes           & \texttimes           & \multicolumn{1}{c|}{Tumor} & 0.859                    & 12.38                  & 1.567                   & 0.698                    & 1.291                    \\
9&& \checkmark           & \texttimes           & \multicolumn{1}{c|}{Organ} & 0.856                    & 12.78                  & 1.635                   & 0.675                    & 1.407                    \\
10& & \setgray \checkmark           & \setgray \checkmark           & \multicolumn{1}{c|}{\setgray Organ} & \setgray 0.857                    & \setgray 12.43                  & \setgray 1.681                   & \setgray 0.691                    & \setgray 1.241                    \\\hline
11&\multirow{5}{*}{TransMorph}            & \checkmark           &     \textendash                 & \multicolumn{1}{c|}{\textendash}      & 0.859                    & 13.35                  & 6.337                   & 1.906                    & 1.785                    \\
12& & \texttimes           & \textendash           & \multicolumn{1}{c|}{\textendash}      & 0.863                    & 13.10                  & 4.923                   & 1.540                    & 1.540                    \\
13& & \texttimes           & \texttimes           & \multicolumn{1}{c|}{Tumor} & 0.858                    & 13.21                  & 4.401                   & 0.877                    & 1.227                    \\
14& & \checkmark           & \texttimes           & \multicolumn{1}{c|}{Organ} & 0.849                    & 13.02                  & 4.651                   & 0.845                   & 1.454                    \\
15&& \setgray \checkmark           & \setgray \checkmark           &  \multicolumn{1}{c|}{\setgray Organ} & \setgray 0.856                    & \setgray 13.19                  & \setgray 4.084                   & \setgray 0.827 & \setgray 1.302     \\ 
\end{tabular}
}
\caption{Effectiveness of adaptive volume-preserving loss in unsupervised registration frameworks on LiTs17 dataset. Three networks, namely VTN in~\cite{zhao2019recursive}, VXM in~\cite{balakrishnan2018unsupervised}, and TransMorph in~\cite{chen2022transmorph}, were applied. The methods used in the study were regular unsupervised (first row), weighted similarity loss only based on the ground truth (GT) mask (second row), volume-preserving loss based on GT(third row), volume-preserving loss on the whole organ based on estimation (fourth row), and on the tumor based on estimation (ours).}
\label{tab:tab2}
\vspace{-0.4cm}
\end{table*}

\subsection{Main Results}

\noindent\textbf{Quantitative Comparison.} 
In Table~\ref{tab:tab1}, we present a comprehensive comparison of our proposed strategy with two traditional methods and three learning-based methods with different network architectures (CNN, cascaded CNN and hybrid transformer-CNN). We compare the methods using a similarity-based manner (regular) following the settings in~\cite{zhao2019recursive}, as well as our volume-preserving manners.
Our results demonstrate that our proposed strategy outperforms all other methods in terms of the STSR metrics, while achieving comparable results in Dice and Landmark Dist, and often better results in folding detection ($|J_\theta|<0$) and smoothness quality ($Std. |J_\theta|$).
Furthermore, we observe that registration networks that achieve better performance in terms of previous metrics such as Dice tend to have worse performance in preserving tumor volumes (STSR). This observation highlights the need to balance these two objectives in order to achieve competitive results.

\noindent\textbf{Qualitative Comparison.} 
Figure~\ref{fig:Fig4} presents a qualitative comparison of our proposed strategy with different state-of-the-art methods.
Comparing the "VTN/VXM/Trans./LapIRN regular" column and the "VTN/VXM/Trans./LapIRN ours" column, it is evident that our method outperforms other methods in terms of volume preservation in the warped image ($1_\text{st}$ row) and the corresponding deformation field ($4_\text{th}$ row).
Furthermore, in terms of visualization of the registration accuracy and the overlay of the tumor ($2_\text{nd}$ row), our proposed method achieves better volume preservation with comparable accuracy compared to other methods.
Lastly, the Jacobian determinant of the deformation field ($3_\text{rd}$ row) reveals that regular methods tend to have a larger Jacobian in tumor regions, indicating greater volume change, compared to our method. This highlights the effectiveness of our volume-preserving approach in preserving tumor volumes while aligning the images.

\subsection{Ablation study}

\noindent\textbf{Volume-Preserving Losses.} In Table~\ref{tab:tab2}, we present the results of our evaluation of the effectiveness of various versions of volume-preserving losses. For each sub-table, the first row (\#1, \#6, and \#11) represents using only the similarity loss, while the second row (\#2, \#7, and \#12) utilizes the ground truth as a weighted similarity loss. The third row (\#3, \#8, and \#13) uses the ground truth binary mask as a tumor mask, while the fourth row (\#4, \#9, and \#14) uses the mask of the entire organ as a tumor mask. The fifth row (\#5, \#10, and \#15) represents our proposed strategy.

By comparing the first and second rows of each sub-table with the fifth row, we observe that our proposed volume-preserving loss can effectively preserve details and achieve superior performance across various metrics. The improvement from the third and fourth rows to the fifth row indicates the importance of adaptiveness in preserving the volume of the organ and tumor, respectively. It should be noted that we do not require pixel-wise ground truth for tumor volume preservation.

\begin{figure}[htbp]
\begin{center}
   \includegraphics[width=0.8\linewidth]{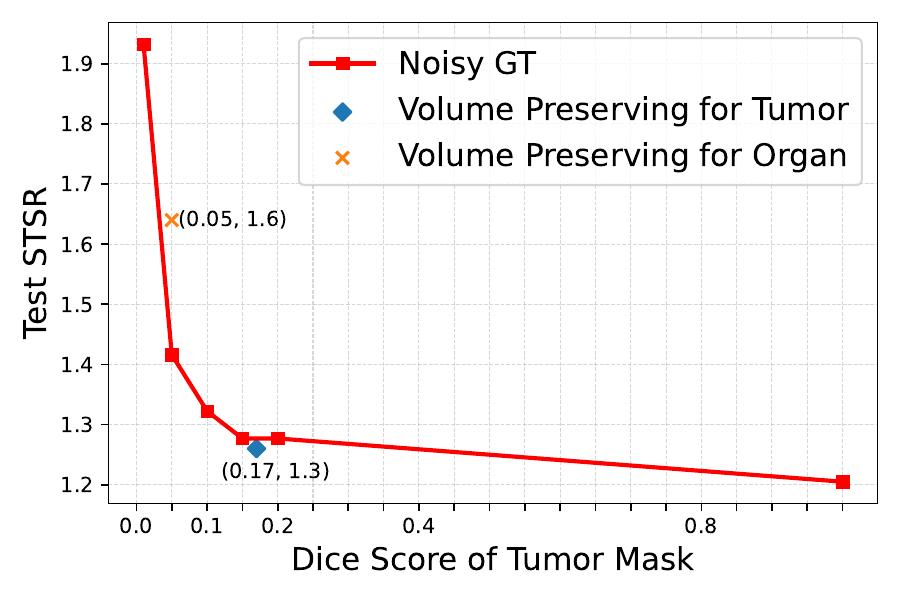}
\end{center}
\vspace{-0.2cm}
   \caption{
   Robustness of our tumor mask estimation. 
   We evaluated the robustness of our tumor mask estimation by running our strategy using RCN on the LiTS datasets with different noisy ground truth (GT) masks as tumor masks, each with specific Dice scores with tumors. We plotted the results on a graph with the Dice score on the x-axis and the STSR in the test set on the y-axis (where smaller values are better).
   }
\label{fig:Fig5}
\vspace{-0.15cm}
\end{figure}

\noindent\textbf{Robustness to Tumor Mask Estimation.} 
In Figure~\ref{fig:Fig5}, we showcase the robustness of our proposed strategy in the presence of noisy ground truth segmentation masks. To evaluate this, we randomly select points within organs to ensure that tumors have a certain Dice score with respect to the ground truth segmentation of tumors.
As depicted in the figure, despite the variations in the noisy ground truth segmentation masks, our strategy's performance remains consistent. Specifically, when the tumor mask has a Dice score above a certain threshold, such as 0.15, the estimated tumor mask consistently achieves a Dice score of approximately 0.17. In contrast, using the whole organ as mask in volume-preserving loss causes degradation in performance as it only has a dice score of roughly 0.05.
This highlights the necessity and efficiency of our tumor mask estimation in stage one.



\section{Conclusions}

In summary, this work addressed the problem of preserving tumor volumes while maximizing similarity during learning-based registration, which is crucial for tracking tumor growth. We proposed a two-stage unsupervised strategy that effectively preserves the size ratio of tumors, without requiring extra networks. Our approach involves generating a soft tumor mask in the first stage using a similarity-based registration network and incorporating an adaptive volume-preserving loss and a weighted similarity loss in the second stage to improve registration performance.
To evaluate our strategy's performance, we introduced a new metric, the Square Tumor Size Ratio (STSR), which measures the preservation of tumor volume. Our proposed strategy was evaluated on various networks and demonstrated superior performance in tumor volume preservation, while achieving comparable results in other metrics. This highlights the effectiveness of our approach in tracking tumor growth and its ability to generalize well to different registration networks. Additionally, we noted that the volume preservation approach could be extended to other lesions besides tumors.\\

\textbf{Acknowledgement}:
This work was supported by the HKSAR Innovation and Technology Commission (ITC) under ITF Project MHP/109/19, the National Natural Science Foundation in China under Grant 62022010, the Beijing Natural Science Foundation Haidian District Joint Fund in China under Grant L222032 and the Fundamental Research Funds for the Central Universities of China from the State Key Laboratory of Software Development Environment in Beihang University in China.

{
\clearpage
\small
\bibliographystyle{iccv2023AuthorKit/ieee_fullname}
\bibliography{iccv2023AuthorKit/egbib}
}

\newpage

\appendix

\begin{center}
\Large\textbf{Supplementary}
\end{center}

This supplementary material includes additional visualizations of brain dataset examples that were not included in the main paper due to space limitations. Furthermore, we visualize landmark locations in warped images between our method and previous approaches on all datasets, as well as an ablation study demonstrating the efficacy and robustness of our proposed method in estimating tumor masks. These additions provide further insight into the effectiveness, robustness and generalization of our volume-preserving methods.

\section{More finaltive Comparisons on Similarity-Based Registration and Volume-Preserving Registration}

\begin{figure*}[htbp]
    \centering
    \includegraphics[trim={0 130 0 150}, clip,width=\linewidth]{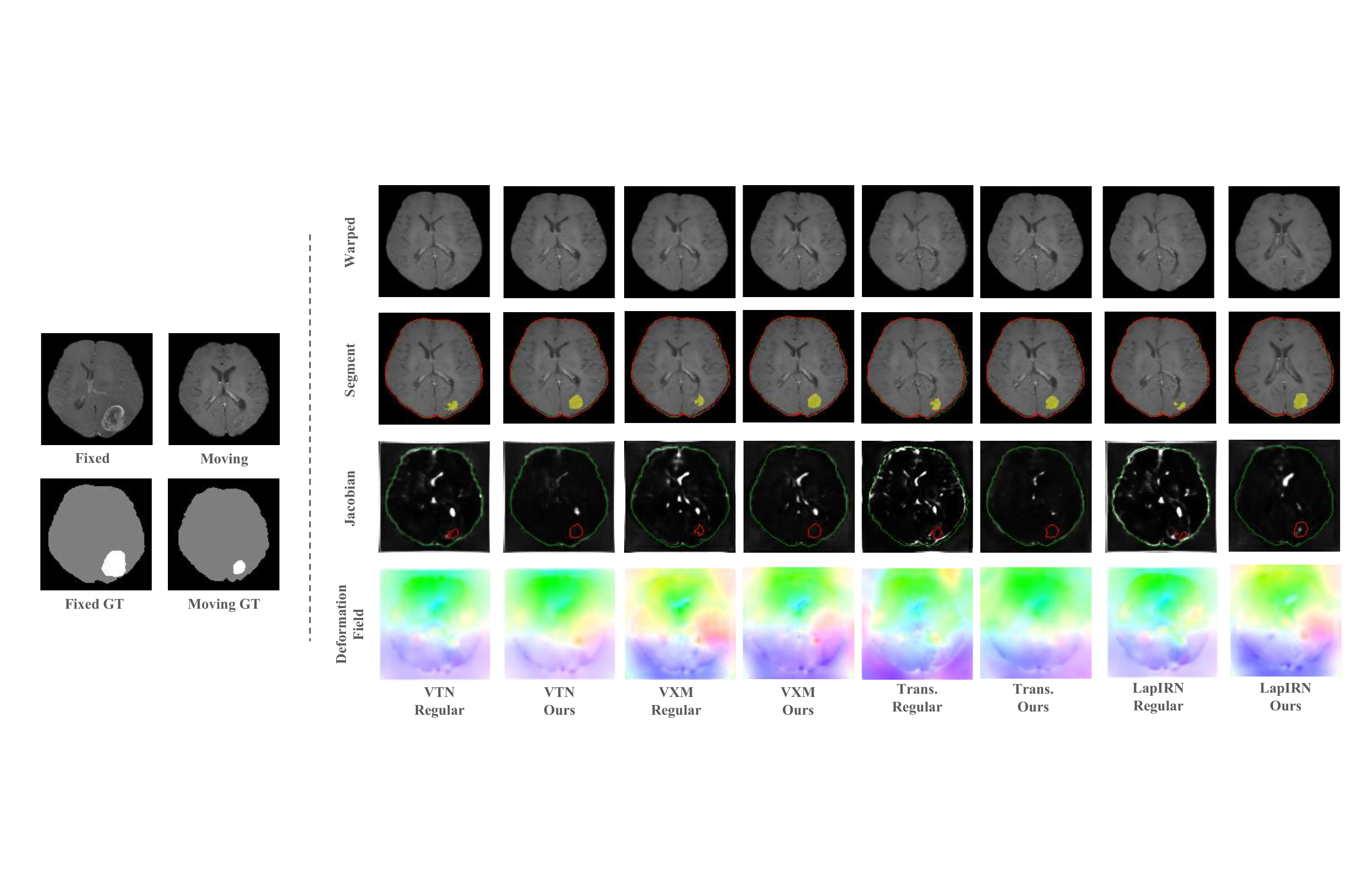}
    \caption{
Qualitative comparison between similarity-based (regular) and volume-preserving (ours) methods trained on the Brain Tumor Segmentation (BraTS20) dataset.
Specifically, the VTN, VXM, and TransMorph (Trans.) networks were tested, both with their regular similarity-based registration versions and our proposed volume-preserving version.
The left side of the figure shows two sets of images: Fixed and Ground Truth (GT), and Moving and GT. The first row of the figure
displays the warped moving image, while the second row illustrates the organ outlines in green and red for the moving and fixed images,
respectively. The yellow overlay highlights the tumors. Our proposed volume-preserving (VP) method ensures the preservation of tumor
volume while aligning the images, as demonstrated by reduced number of visible changes in tumor size. In the third row, the Jacobian
matrix of the deformation field is visualized. The green and red lines represent the organ and tumor outlines, respectively. The white
areas indicate a large Jacobian, which corresponds to a more significant change in volume. 
The method without volume-preserving loss demonstrates a larger white area in the tumor, indicating a greater volume change of tumor volume. The last row of the figure displays the
deformation field.
}
    \label{fig: main-brain}
\end{figure*}

In addition to the results presented in Figure 4. of the main paper, we provide further qualitative results on brain scans using three different networks (VTN, VXM and TransMorph), as shown in Figure~\ref{fig: main-brain}. These additional results provide further evidence of the effectiveness of our volume-preserving approach in preserving tumor volumes. 

\begin{figure*}[htb]
    \centering
    \includegraphics[width=\linewidth]{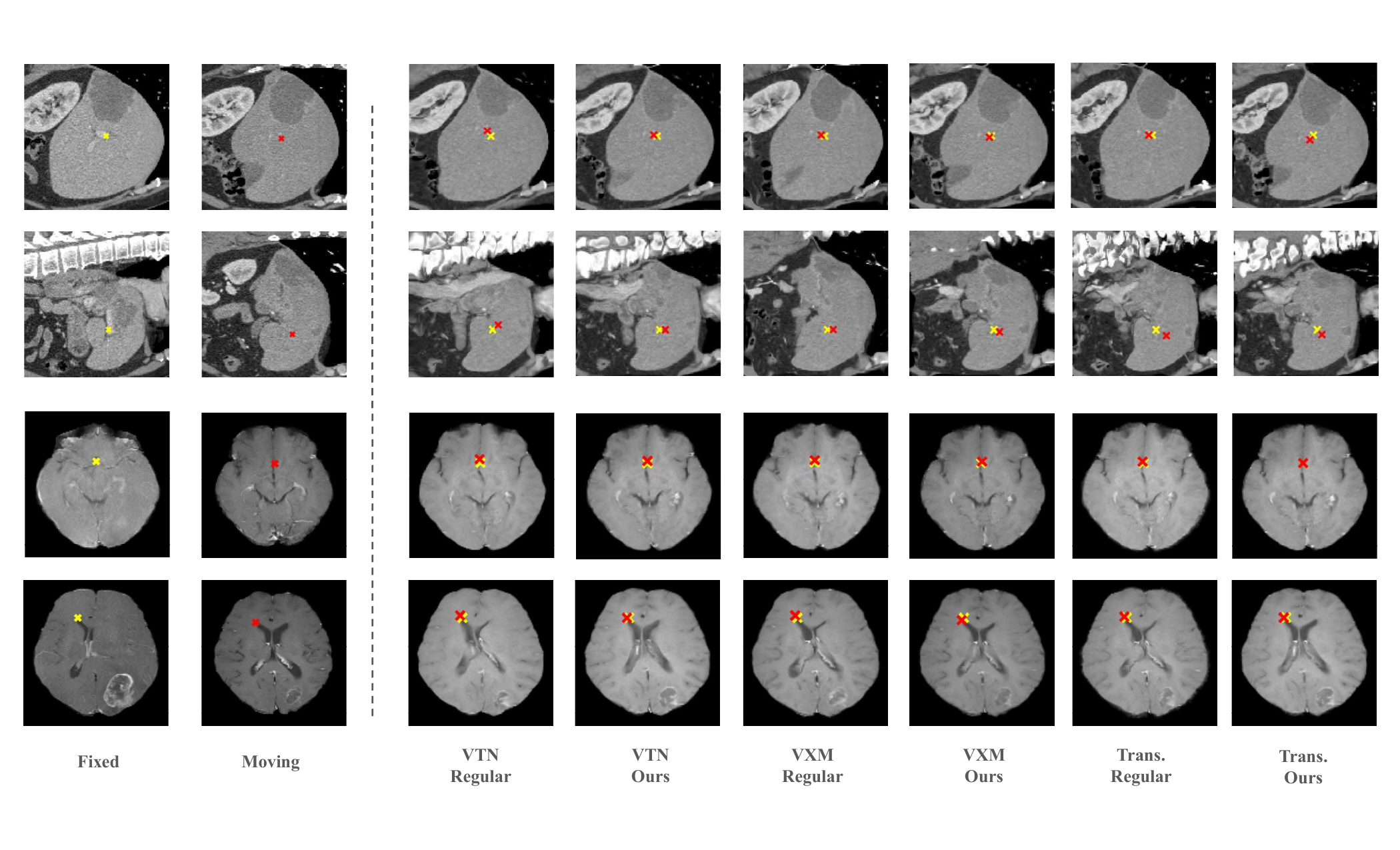}
    \caption{
Comparison of similarity-based (regular) and volume-preserving (ours) methods using landmark visualization trained on the  Liver Tumor Segmentation (LiTS17) dataset (1st and 2nd row) and the Brain Tumor Segmentation (BraTS20) dataset (3rd and 4th row). Specifically, the VTN, VXM, and TransMorph (Trans.) networks were tested, both with their regular similarity-based registration versions and our proposed volume-preserving version. 
The left side of the figure comprises two sets of images, namely, Fixed and its corresponding landmark locations represented by a yellow cross and Moving and its landmark location represented by a red cross. The right side of the figure shows the warped image using different methods, along with the warped landmarks from the moving image represented by a red cross and the corresponding landmarks from the fixed image projected on the plane represented by a yellow cross.
The landmark visualization results indicate that our proposed method yields landmark distances that are comparable to those obtained using similarity-based methods while simultaneously preserving tumor volume. This aspect is highly beneficial in tracking tumor growth.
    }
    \label{fig: lilmk}
\end{figure*}

Furthermore, we visualize landmark locations in the warped images for the regular similarity-based method and our proposed method, as depicted in Figure~\ref{fig: lilmk}. These visualizations demonstrate the ability of our approach to align the anatomy in the images comparably to previous methods on different networks.

\section{Ablation Study on Calculation of Soft Tumor Mask}

Table~\ref{tab:tab3} presents the results of various calculation methods used to estimate tumor masks on the LiTs17 dataset with VTN. The study demonstrates the adaptive methods that generate soft masks for tumor estimation (1st and 2nd row)  produce similar results, while the use of hard threshold functions that generates binary masks (3rd, 4th, and 5th row) fails to balance the preservation of the tumor size ratio (STSR) with the alignment of image anatomy (Dice and Landmark Distance (Lm. Dist)).
Moreover, the study shows that the adaptive volume-preserving loss is robust for different transformation functions. Two different transformation functions in adaptive methods , sigmoid (1st row) and sin (2nd row), achieve comparable performance on all three metrics.

In practice, the complete transformation function for "Sigm" is expressed as $STM(x) = sigmoid(5\cdot (D(x)-1.5))$, while for "Sin" it is given by $STM(x) = {1\over 2}sin(\pi\cdot (D(x)-1.5))+0.5$, where $D$ denotes the size ratio change of the voxel at location $x$. The exact forms have been chosen to ensure that $STM(x)\approx 1$ when $D(x)\geq 2$ and $STM(x)\approx 0$ when $D(x)=1$. This is because we observed when the change of size ratio exceeds 2, it is highly probable that tumors are present. Conversely, if the ratio does not deviate significantly (changes close to 1), it is more likely that the observed tissue is normal.

These results provide valuable insights into the optimization of tumor estimation methods and suggest the use of adaptive methods in volume-preserving loss to enhance performance.

\begin{table}[th]
\centering
\begin{tabular}{ll|ccc}
\hline
Trans. & Thres. & Dice $\uparrow$ & Lm. Dist $\downarrow$ & STSR $\downarrow$ \\ 
\hline
Sigm $\dagger$  &       & \textbf{0.908}     &   10.89       &  \textbf{1.26}   \\
Sin    &       &  0.906    &  \textbf{10.77}      &   1.29  \\
Hard   & 1      &  0.840    &   12.23       &  1.27   \\
Hard   & 1.5    & 0.877    &   13.67       &  1.46   \\
Hard   & 2      &  0.904    &   12.68       &  1.86  
\end{tabular}
\caption{
Comparison between different calculations of tumor mask on VTN network using LiTS17 dataset. 
The term "Trans." refers to the transformation function that computes the estimated tumor masks based on the change of size ratio in each voxel. It encompasses various functions, such as sigmoid ("Sigm"), sin ("Sin"), which predict soft tumor masks, and threshold functions ("Hard"), that generate binary masks based on a fixed threshold.
"Thres." refers to the threshold for binary tumor mask estimation. 
The "$\dagger$" indicates the adoption of this transformation function used in previous experiments.}
\label{tab:tab3}
\end{table}

\end{document}